\documentstyle[12pt]{article}

\makeatletter
\@addtoreset{equation}{section}
\makeatother

\newcommand{\bz}{{\hbox{\bf Z}}}
\newcommand{\br}{{\hbox{\bf R}}}
\newcommand{\bc}{{\hbox{\bf C}}}

\def\Matrix#1#2#3#4#5{
\dimen6=#5
\dimen7=\dimen6
\dimen10=\dimen6
\advance \dimen10 by 6pt
\divide \dimen6 by 2
\dimen8=\dimen6
\dimen9=\dimen6
\advance \dimen8 by 0.2pt
\advance \dimen9 by -0.2pt
#1\kern-1mm
\raise1.8pt\hbox{$
{\mathop{\hbox to \dimen10{
\vrule width \dimen7 depth0pt height.4pt
\hskip-\dimen8
\vrule height \dimen6 depth \dimen6 width.4pt
\hskip \dimen9
\hfill}}
\limits^{\textstyle #2}_{\textstyle #3}}$}
\kern-.7mm #4
}

\def\bw(#1,#2,#3,#4){\Matrix{#1}{#2}{#3}{#4}{0.8cm}}

\begin{document}

\begin{flushright}
  UT-603 \\
  (Revised Version) \\
  Aug. 1992
\end{flushright}
\vspace{24pt}
\begin{center}
\begin{Large}
{\bf Yang-Baxter Equation for
$A^{(1)}_{n-1}$ Broken ${\Bf Z}_N $ Models}
\end{Large}

\vspace{36pt}

Yas-Hiro Quano \raisebox{2mm}{$\star$}

\vspace{6pt}
{\it Department of Physics, University of Tokyo,} \\
{\it Bunkyo-ku, Tokyo 113, Japan}

and \\
Akira Fujii \raisebox{2mm}{$\star\star$}

\vspace{6pt}
{\it Institute of Physics, College of Arts and Sciences,} \\
{\it University of Tokyo, Meguro-ku, Tokyo 153, Japan}

\underline{ABSTRACT}
\end{center}

\vspace{48pt}

We construct a factorized representation of
the $\frak g \frak l _n$-Sklyanin algebra
from the vertex-face correspondence.
Using this representation,
we obtain a new solvable model which gives an
$\frak s \frak l _n$-generalization of the broken $\bz _N $ model.
We further prove the Yang-Baxter equation for this model.
\end{center}

\vspace{24pt}

\vfill
\hrule

\vskip 3mm
\begin{small}

\noindent\raisebox{2mm}{$\star$}
A Fellow of the Japan
Society of the Promotion
of Science for Japanese Junior Scientists.

\noindent Partly supported by the Grant-in-Aid for Scientific Research
from the Ministry of Education, Science and Culture
(No. 04-2297).

\noindent\raisebox{2mm}{$\star\star$}
A Fellow of the Japan
Society of the Promotion
of Science for Japanese Junior Scientists.

\noindent Partly supported by the Grant-in-Aid for Scientific Research
from the Ministry of Education, Science and Culture
(No. 04-2747).

\end{small}

\begin{document}
\maketitle

\newpage
\section{Introduction}
The Yang-Baxter equation (YBE) has been studied as the master equation
of solvable lattice statistical models in two dimensions \cite{YBE}
for over two decades.
The problem of classifying or enumerating solutions to YBE
is one of the central subjects in mathematical physics.
Infinite dimensional symmetries play a remarkable role
to investigate exactly solvable models
as the quantum affine algebra \cite{Hopf} provides a prescription of
systematic construction of the trigonometric solutions to YBE.
For the elliptic solutions, however,
no counterpart of the quantum affine algebra has yet been known.
It is urgently necessary to find such symmetries
incarnated in the elliptic solution to YBE.

The Sklyanin algebra \cite{S1} and its generalizations
\cite{Sn}\cite{HW}\cite{QF} have been recognized to be useful
to such a problem.
The Sklyanin algebra appears naturally with the $L$-operator
in the context of the quantum inverse method \cite{QIM}.

Frenkel and Reshetikhin \cite{FR} showed that
the correlation functions of the $q$-deformation of the vertex operators
satisfy $q$-difference equations (see also \cite{DFJMN}\cite{KQS}),
and that the resulting connection matrices give the elliptic solution
of YBE \cite{ABF}\cite{ABCD}.
They introduced the $L$-operator
as generators of the quantum affine algebra.
Since the currents generate the affine Lie algebra,
it is very likely that the $q$-analog of the current algebra is
the Sklyanin algebra.

Furthermore,
the representation theory of the Sklyanin algebra
is rich enough to derive some elliptic models \cite{HY}\cite{QF2}\cite{Ser}.
Hasegawa and Yamada \cite{HY} observed
that the $R$-matrix of the broken $\bz _N$ model \cite{KM}\cite{1P}
can be obtained
from that of the eight vertex model \cite{ESM},
using the cyclic representation of
the $\frak g \frak l _2$-Sklyanin algebra \cite{S1}.
Through analysis of their paper
we found out that the vertex-face correspondence \cite{PQ}
is the key to open the door leading to the new world.

In the present paper
we show that the $L$-operator for
the $\bz _n \otimes \bz _n $-symmetric model \cite{Zn}
factorizes elementwise into two pieces,
which are essentially
the left and right intertwining vectors \cite{JMO}.
We also construct the $A^{(1)}_{n-1}$ broken $\bz _N$ model
by utilizing the ``factorized representation''
of the Sklyanin algebra.
Finally we prove YBE for our new model.

Our strategy is as follows:

$1^{\circ }$: Prepare an elliptic solution to YBE
\begin{equation}
\check{R}^{(23)}(z_1 -z_2)\check{R}^{(12)}(z_1)\check{R}^{(23)}(z_2)=
\check{R}^{(12)}(z_2)\check{R}^{(23)}(z_1)\check{R}^{(12)}(z_1 -z_2).
\end{equation}
Here we use Belavin's solution of
the $\bz _n \otimes \bz _n $-symmetric model \cite{Zn}.

$2^{\circ }$: Solve the defining equation of the Sklyanin algebra
\begin{equation}
\check{R}(z_1 -z_2)L(z_1)\otimes L(z_2)=
L(z_2)\otimes L(z_1)\check{R}(z_1 -z_2).
\end{equation}
This equation is the same as that appearing in
the quantum inverse problem \cite{QIM}.

$3^{\circ }$: Solve the following equation
\begin{equation}
S(z_1-z_2 )L(z_1 )\otimes L(z_2 )=
L(z_2 )\otimes L(z_1 )S(z_1-z_2 ).
\end{equation}
A solution $S(z)$ define a new model.

$4^{\circ }$: Prove YBE for $S(z_1 ,z_2 )$
\begin{equation}
S^{(23)}(z_1-z_2 )S^{(12)}(z_1 )S^{(23)}(z_2 )
=S^{(12)}(z_2 )S^{(23)}(z_1 )S^{(12)}(z_1 -z_2 ).
\end{equation}

{}~

Let us explain the contents of this paper.
In section 2
to establish notation and for later convenience,
we summarize those results
needed in the subsequent sections,
concerning the $\bz _n \otimes \bz _n$ symmetric model \cite{Zn},
the cyclic $A^{(1)}_{n-1}$ model \cite{ABCD}\cite{Kun}, and
the vertex-face correspondence \cite{JMO} between these two models.
In section 3 we give a brief review for
the $\frak g \frak l _n$-Sklyanin algebra,
which is denoted by $\frak F _n$ in the main text.
Furthermore,
we construct a ``factorized representation''
of $\frak F _n$ \cite{INS}
on the basis of the vertex-face correspondence.
In section 4 we derive an $\frak s \frak l _n$-generalization of
the broken $\bz _N $ model.
The number of independent local states of this model is $N^{n-1}$.
We also prove YBE
for this model.
In section 5 we summarize our results and give some remark.
In appendix A a simple case study for $n=3, N=2$ is given.

{}~

\section{Intertwining Vector}
2.1~~$\bz _n \otimes \bz _n$-{\it Symmetric Model} ~~~~~~
To begin with, let us fix notation concerning
the theta functions.
For a complex number $\tau $ in the upper half-plane,
let $\Lambda _{\tau }:=\{ \xi _1 \tau +\xi _2 | \xi _1 , \xi _1 \in \bz \}$
be the lattice generated by $1$ and $\tau$,
and $E_{\tau }:= \bc /\Lambda _{\tau }$ the complex torus
which can be identified with an elliptic curve.
The Riemann's theta function
with the characteristics $a,b\in \br $ is defined by the
following convergent series \cite{Tata}
\begin{equation}
\vartheta\left[\begin{array}{c} a \\ b \end{array} \right](z,\tau ):=
\sum_{m\in \bz }
\exp \left\{ \pi \sqrt{-1}(m+a)~\left[ (m+a)\tau
+2(z+b) \right] \right\}.
\end{equation}
The zeros are
\begin{equation}
\vartheta\left[\begin{array}{c} a \\ b \end{array} \right](u,\tau )=0~~
\mbox{at~~} u=(\frac{1}{2}-a)\tau +(\frac{1}{2}-b),~~~~\mbox{mod~~}
\Lambda _{\tau }. \label{zero}
\end{equation}

Hereafter a positive integer $n \geq 2$ is fixed and
we will use the following compact symbols
\begin{equation}
\sigma_{\vec{\alpha }}(z)=
\vartheta\left[\begin{array}{c} \alpha _2 /n +1/2 \\ \alpha _1 /n+1/2
\end{array} \right](z,\tau ), ~~~~~~
\theta^{(j)}(z)=
\vartheta\left[\begin{array}{c} 1/2 - j/n \\ 1/2 \end{array}
\right](z,n\tau ),
\end{equation}
for $\vec{\alpha }=(\alpha _1,\alpha _2)\in \bz \otimes \bz$ and
for $j\in \bz _n $.

Now we consider the vertex model on a two-dimensional
lattice ${\cal L}$ such that the state variables take on
$\bz _n $-spin.
Set $\check{R}_{ik}^{jl}$ be a Boltzmann weight
for a single vertex with bond states
$i, j, k, l \in \bz _n$
\begin{equation}
\check{R}_{ik}^{jl}=\bw(j,i,l,k) .
\end{equation}
Let $V=\bc ^n $ and $\{\epsilon _i \}_{i\in \bz _n }$ be
the standard orthonormal basis of $V$.
The Boltzmann weights give a representation of
the linear map $\check{R}: V\otimes V \rightarrow V\otimes V$
on the basis $\{ \epsilon _j \otimes \epsilon _l \}$
\begin{equation}
\check{R}(\epsilon _j \otimes \epsilon_ l )=\sum_{i, k \in \bz _n }
(\epsilon _i \otimes \epsilon _k )
\check{R}_{ik}^{jl}.
\end{equation}
$\check{R}$ is also called the $R$-matrix.

Belavin \cite{Zn} considered the $\bz _n \otimes \bz _n $-symmetric model
whose  $R$-matrix satisfies the following properties
\begin{eqnarray}
&&({\romannumeral 1})~~~~\check{R}_{ik}^{jl}=0,
\mbox{~~unless $i+k=j+l$,~~mod $n$}, \\
&&({\romannumeral 2})~~~~\check{R}_{i+p,k+p}^{j+p,l+p}=
\check{R}_{ik}^{jl},
\mbox{~~for every $i,j,k,l,$ and $p\in \bz _n$}.
\end{eqnarray}

Let $g$ and $h$ be the linear maps in $V$ defined by
\begin{equation}
g\epsilon _i =\omega ^i \epsilon _i ,~~~~~~
h\epsilon _i =\epsilon _{i-1},
\end{equation}
where, $\omega =\exp (2\pi \sqrt{-1}/n).$
Matrices $I_{\vec{\alpha }}=g^{\alpha _1}h^{\alpha _2}$,
for $\vec{\alpha }=(\alpha _1,\alpha _2)\in \bz _n \otimes \bz _n$,
span a basis of $\frak g \frak l _n$.
Then $R$-matrix of Belavin's model,
which is a function of the spectral parameter
$z\in E_{\tau }$, is expressed as follows

\begin{equation}
\check{R}(z)=P\sum_{\vec{\alpha }\in \bz _n \otimes \bz _n}
u_{\vec{\alpha }}(z)I_{\vec{\alpha }} \otimes I_{\vec{\alpha }}^{-1}.
\label{R1}
\end{equation}
Here, $P$ denotes the transposition matrix on $V\otimes V$
\begin{equation}
P(x\otimes y)=y\otimes x,
\end{equation}
and
\begin{equation}
u_{\vec{\alpha }}(z):=\frac{1}{n}\frac
{\sigma_{\vec{\alpha }}(z+w/n)}{\sigma_{\vec{\alpha }}(w/n)}, \label{R2}
\end{equation}
where  $w\neq 0~\mbox{mod~}\Lambda _{\tau }$, is a constant.
Belavin \cite{Zn} conjectured and some authors \cite{CBT} proved that
the matrix $\check{R}(z)$ solves YBE:

{}~

{\bf Theorem 2.1} \cite{CBT}
\begin{equation}
\check{R}^{(23)}(z_1 -z_2)\check{R}^{(12)}(z_1)\check{R}^{(23)}(z_2)=
\check{R}^{(12)}(z_2)\check{R}^{(23)}(z_1)\check{R}^{(12)}(z_1 -z_2),
\label{YB}
\end{equation}
{\it where $\check{R}^{(12)}(z)=\check{R}(z)\otimes I$
and $\check{R}^{(23)}(z)=I\otimes \check{R}(z)$
are matrices acting on }
$V^{\otimes 3}$.

{}~

2.2~~{\it Cyclic $A^{(1)}_{n-1}$ model}~~~~~~
Here we fix the notation of affine Lie algebra
$A^{(1)}_{n-1}$ \cite{KMA}.
Let $\epsilon _j (0\leq j<n)$ be the same ones
introduced in the previous subsection.
The inner product $(~|~)$ in $V=\bc ^n $ is the standard one:
$(\epsilon _j |\epsilon _k )=\delta _{jk}$.
The fundamental weights $\Lambda _j (0\leq j<n)$ are defined by
\begin{equation}
\bar{\epsilon}_j =\epsilon _j -\epsilon =\Lambda _{j+1} -\Lambda _{j},
{}~~~~\epsilon =\frac{1}{n}\sum_{j=0}^{n-1} \epsilon _j,
\end{equation}
where, $\Lambda _n =\Lambda _0.$
We set $\rho =\displaystyle \sum_{j=0}^{n-1} \Lambda _j $
and $\delta$ denotes the null root.

Let $\frak h :=\bc \Lambda _0 \oplus \cdots \bc \Lambda _{n-1}
\oplus \bc \delta $ be a $\bc $-vector space
and the inner product on $\frak h $ is induced from that of $V$ and
$(\Lambda _i |\delta )=1,$
$(\delta |\delta )=0.$
For $a\in \frak h $,
$(a|\delta )$ signifies the level of $a$.
An ordered pair $(a,b)~ (a,b\in \frak h )$ is called admissible if
\begin{equation}
b=a+\bar{\epsilon}_j , ~~\mbox{for some $j\in \bz _n.$}
\end{equation}
We introduce ${\cal A}:=\bz \bar{\epsilon } _0
+ \cdots + \bz \bar{\epsilon } _{n-1}$ and set
\begin{equation}
\frak h (a_0 ):=\{ a|a-a_0 \in {\cal A} \},
\end{equation}
for $a_0 \in \frak h $.
We call $P=\bz \Lambda _0 \oplus \cdots \bz \Lambda _{n-1}
\oplus \bz \delta \subset \frak h $
the weight lattice of $A^{(1)}_{n-1}$.
We put
\begin{equation}
P_{cl}=\bz \Lambda _0 \oplus \cdots \oplus \bz \Lambda _{n-1}, ~~~~
P_{cl}^{+}=\bz _{\geq 0}\Lambda _0 \oplus \cdots
\oplus \bz _{\geq 0}\Lambda _{n-1}.
\end{equation}
We also put
\begin{equation}
P_N =\{ a\in P_{cl}| ~(a|\delta )=N \}, ~~~~
P_N ^{+}=\{ a\in P_{cl}^{+}| ~(a|\delta )=N \}.
\end{equation}

Fix $a_0 \in \frak h $
and consider the IRF model \cite{ESM} on a two dimensional square lattice
${\cal L}^* $,
which is a dual lattice of ${\cal L}$ considered in subsection 2.1,
such that
the state variables take on values of $\frak h (a_0 )$.
Let
$\displaystyle W
\left( \begin{array}{cc} a & b \\ d & c \end{array} \right) $
be a Boltzmann weight for a state configuration
$\displaystyle \left( \begin{array}{cc} a & b \\ d & c \end{array} \right) $
round a face.
Here the four states $a, b, c, d$ are ordered clockwise
from the S corner. (See Figure 1.)
We set $W
\left( \begin{array}{cc} a & b \\ d & c \end{array} \right) =0~~$
unless the four pairs $(a,b), (a,d), (b,c)$, and $(d,c)$ are admissible.
Non-zero Boltzmann weights are parametrized in terms of
the elliptic theta function of the spectral parameter $z$
as follows:
\begin{equation}
\begin{array}{rcl}
W_{z}
\left( \begin{array}{cc} a & a+\bar{\epsilon }_j \\
a+\bar{\epsilon }_j & a + 2 \bar{\epsilon }_j \end{array} \right) & = &
\displaystyle\frac{h(z+w)}{h(w)}, \\
{}~ & ~ & ~ \\
W_{z}
\left( \begin{array}{cc} a & a+\bar{\epsilon }_j \\
a+\bar{\epsilon }_j & a+\bar{\epsilon }_j +\bar{\epsilon }_k
\end{array} \right) & = &
\displaystyle\frac{h(a_{jk}w-z)}{h(a_{jk}w)} ~~~~(j\neq k), \label{BW} \\
{}~ & ~ & ~ \\
W_{z}
\left( \begin{array}{cc} a & a+\bar{\epsilon }_k \\
a+\bar{\epsilon }_j & a+\bar{\epsilon }_j +\bar{\epsilon }_k
\end{array} \right) & = &
\displaystyle\frac{h(z)}{h(w)}
\frac{\sqrt{h(a_{jk}w+w)}\sqrt{h(a_{jk}w-w)}}{h(a_{jk}w)} ~~~~(j\neq k).
\end{array}
\end{equation}
Here $h(z)$ is defined by
\begin{equation}
h(z):=\prod_{j=0}^{n-1} \theta^{(j)}(z)/
\prod_{j=1}^{n-1} \theta^{(j)}(0), \label{hz}
\end{equation}
$w$ is a complex constant, and
\begin{equation}
a_{jk}=\bar{a}_j -\bar{a}_k;~~~
\bar{a}_j =(a+\rho | \epsilon _j )+\frac{2i+1-n}{2n}.
\end{equation}
Here we use a different definition of $\bar{a}_j $ from that of \cite{ABCD}.
In \cite{ABCD} $\bar{a}_j =(a+\rho | \epsilon _j )$.
This modification is essential
if and only if $w$ is a rational number.
Let $N$ be a positive integer.
In the context of \cite{ABCD},
we can restrict the local states to the elements of
$P_N ^{+}$ when $w=1/(N+n)$.
In our definition of $\bar{a}_j $, if we choose $w=1/N$
we have the periodicity of the period $N$
\begin{equation}
W_{z} \left( \begin{array}{cc}
a+N\bar{\epsilon }_j & b+N\bar{\epsilon }_j \\
d+N\bar{\epsilon }_j & c+N\bar{\epsilon }_j \end{array} \right) =
W_{z} \left( \begin{array}{cc} a & b \\ d & c \end{array} \right) .
\end{equation}
Kuniba \cite{Kun} and Deguchi \cite{De}
studied other $A^{(1)}_{n-1}$ models
with the periodicity.
We call the model defined by (\ref{BW}) with $w=1/N$
the cyclic $A^{(1)}_{n-1}$ model.

{}~

2.3~~{\it Vertex-Face Correspondence}~~~~~~
Jimbo, Miwa and Okado \cite{JMO}
introduced the following intertwining vectors
to show the equivalence between
the $\bz _n \otimes \bz _n$-symmetric model and
the $A^{(1)}_{n-1}$ model.
In addition to the difference of the definition of $\bar{a}_j $,
the Boltzmann weights in \cite{ABCD} and ours
differ by the freedom of the gauge transformation.
Hence we must use the following intertwining vector (see Figure 2a)

\begin{equation}\begin{array}{rl}
\phi _{ab}(z):=  &
{}~^{t}(\phi_{ab}^{(0)}(z), \cdots , \phi_{ab}^{(n-1)}(z)), \\
\phi _{ab}^{(i)}(z):= & \left\{ \begin{array}{ll}
\theta^{(i)}(z-nw\bar{a}_j )/\sqrt{\Delta _j (a)} ~~ &
\mbox{if $b=a+\bar{\epsilon}_j ,$} \\
0 ~~ & \mbox{otherwise,} \end{array} \right. \end{array}
\end{equation}
where,
\begin{equation}
\Delta _j (a):=\prod _{l=0}^{j-1} h(a_{lj}w)
\prod _{l=j+1}^{n-1} h(a_{jl}w).
\end{equation}
We then have (see Figure 2b)

{}~

{\bf Theorem 2.2} \cite{JMO}
\begin{equation}
\check{R}(z_1 -z_2 ) \phi _{ad}(z_1 )\otimes \phi _{dc}(z_2 )=
\sum_{b} W_{z_1 -z_2 }
\left( \begin{array}{cc} a & b \\ d & c \end{array} \right)
\phi _{ab}(z_2 )\otimes \phi _{bc}(z_1 ). \label{JMO}
\end{equation}

{}~

We call $\phi _{ab}(z)$ the left intertwining vector.
The following important corollary \cite{JMO}
can be immediately derived from Theorems 2.1 and 2.2.

{}~

{\bf Corollary 2.3}~~~~{\it Boltzmann weights (\ref{BW})
satisfy YBE for IRF-type }
\begin{equation}
\begin{array}{c}
\displaystyle \sum_{g}
W_{z_1 -z_2}\left( \begin{array}{cc} a & b \\ g & c \end{array} \right)
W_{z_1 }\left( \begin{array}{cc} g & c \\ e & d \end{array} \right)
W_{z_2 }\left( \begin{array}{cc} a & g \\ f & e \end{array} \right) \\
{}~ \\
\displaystyle =\sum_{g}
W_{z_2 }\left( \begin{array}{cc} b & c \\ g & d \end{array} \right)
W_{z_1 }\left( \begin{array}{cc} a & b \\ f & g \end{array} \right)
W_{z_1 -z_2}\left( \begin{array}{cc} f & g \\ e & d \end{array} \right).
\end{array}
\end{equation}

{}~

\section{Sklyanin Algebra and Its Representation}
3.1~~{\it Sklyanin Algebra}~~~~~~
Let us consider the $L$-operator satisfying
\begin{equation}
\check{R}(z_1 -z_2)L(z_1)\otimes L(z_2)=
L(z_2)\otimes L(z_1)\check{R}(z_1 -z_2), \label{QIM}
\end{equation}
which is the defining equation for
the $\frak g \frak l _n$-Sklyanin algebra \cite{Sn}\cite{QF}.

Here we summarize the results obtained in \cite{QF}.
Assume that $L(z)$ has the form
\begin{equation}
L(z)=\sum_{\vec{\alpha }\in \bz _n \otimes \bz _n}
u_{\vec{\alpha }}(z)I_{\vec{\alpha }} S^{\vec{\alpha }}. \label{RLL}
\end{equation}
Note that
$L(z)$ is an operator-valued matrix;
$S^{\vec{\alpha }}$'s denote operators
acting on a quantum space
while $I_{\vec{\alpha }}$'s act on an auxiliary space $V=\bc ^n$.

Substituting (\ref{R1}) and (\ref{RLL}) into (\ref{QIM}) we obtain

{}~

{\bf Theorem 3.1}\cite{QF}~~~~
{\it For each $\vec{a},\vec{b}\in \bz _n\otimes \bz _n$, }
\begin{equation}
\sum_{\vec{\alpha }\in \bz _n\otimes \bz _n}
J_{\vec{a}\vec{b}}^{\vec{\alpha }}(w)
\left[ S^{\vec{a}-\vec{\alpha }}, S^{\vec{b}+\vec{\alpha }}\right]
_{(\vec{\alpha}, \vec{a}-\vec{b})}=0. \label{STR}
\end{equation}
{\it Here, ``$\omega$-commutators'' are defined by }
\begin{equation}
[A,B]_{(\vec{x},\vec{y})}:=\omega ^{x_2 y_1}AB-\omega ^{x_1 y_2}BA,
\end{equation}
{\it and the structure constants }
{\it $J_{\vec{a}\vec{b}}^{\vec{\alpha }}(w)$'s are given by }
\begin{equation}
J_{\vec{a}\vec{b}}^{\vec{\alpha }}(w)
=\frac{\omega ^{-\alpha _2 (\alpha _1+1)}
\sigma_{2\vec{\alpha } -\vec{a}+\vec{b}}(0)}
{(\sigma_{\vec{\alpha }}\sigma_{\vec{a}-\vec{\alpha }}
\sigma_{\vec{a}-\vec{b}-\vec{\alpha }}
\sigma_{\vec{b}+\vec{\alpha }})(w/n)}.
\end{equation}

{}~

The ``$\omega$-commutation'' relations (\ref{STR})
generate a two-sided ideal
$\frak A$ in the free associative algebra $A$
generated by
$\{S^{\vec{\alpha }}\}_{\vec{\alpha }\in \bz _n \otimes \bz _n }$.

{}~

{\bf Definition 3.2}~~\cite{QF}~~~~
{\it The quotient algebra $\frak F _n:=A/\frak A$ is called
the $\frak g \frak l _n $-Sklyanin algebra.}

{}~

This object is a quadratic generalization of
$U(\frak g \frak l _n )$, the universal enveloping algebra
of the Lie algebra $\frak g \frak l _n$.
Obviously setting $S^{\vec{a}}=I_{\vec{a}}^{-1}$ (hence $L(z)=R(z)$)
gives a representation of the $\frak F _n $.
In this case $\frak F _n $ is $U(\frak g \frak l _n )$ itself.
We notice in the preceding paper \cite{QF2}
that a certain good representation of
$\frak F _n $ is relevant to obtain a generalization
of the broken $\bz _N $ model \cite{KM}.
We construct a ``factorized representation'' of $\frak F _n $
in the next subsection.

{}~

3.2~~{\it Factorized Representation of the Sklyanin Algebra}~~~~~~
Now we construct a nontrivial representation of $\frak F _n$.
In this representation
$L$-operator is factorized to a product of
the left and right intertwining vector of
the $\bz _n \otimes \bz _n $-symmetric model
and the unrestricted $A^{(1)}_{n-1}$ model.

We regard
$\phi_{aa+\bar{\epsilon}_j }^{(i)}(z)$ as an $(i,j)$-component
of a function-valued matrix $\Phi _a (z)$.
Let $\tilde{\Phi }_a (z)$ be a cofactor matrix of $\Phi _a (z)$ and set
(see Figure 3a)
\begin{equation}
\begin{array}{rl}
\bar{\phi }_{ab}(z):= &
(\bar{\phi }_{ab}^{(0)}(z), \cdots , \bar{\phi }_{ab}^{(n-1)}(z)), \\
\bar{\phi }_{aa+\bar{\epsilon }_j }^{(i)}(z):= &
(\tilde{\Phi }(z))_{j}^{i}/Ch(z-(n-1)/2), \label{phibar}
\end{array}
\end{equation}
where we assume $z\neq (n-1)/2$ and
\begin{equation}
C=e^{\pi\sqrt{-1}[(n-1)(n+2)+2(n-1)^2 \tau ]/2n}
\displaystyle\frac{
\prod_{i=0}^{n-1} (\sum_{j=0}^{n-1} \omega^{ij} \theta ^{(j)}(1/2))}
{h(n/2-(n-1)\tau /2)\prod_{k=1}^{n-1} h(k\tau /n)^{n-k}}, \label{C}
\end{equation}
is a constant.

Note that $\phi_{ab}(z)$ is a column vector while
$\psi_{ab}(z)$ is a row vector.
Then by the rule of multiplication of matrices in $V=\bc ^n $,
$\bar{\phi }_{ab}(z)\phi_{cd}(z)$ is a scalar function and
$\phi_{ab}(z)\bar{\phi }_{cd}(z)$ is a function-valued matrix.
It is clear that $\phi_{ab}(z)$ and $\bar{\phi }_{ab}(z)$
enjoy the following orthogonal properties

{}~

{\bf Proposition 3.3}
\begin{eqnarray}
\bar{\phi }_{aa+\bar{\epsilon }_i }(z)\phi_{aa+\bar{\epsilon }_k }(z) & = &
\delta_{ik}, \\
\sum_{j=0}^{n-1}
\phi_{aa+\bar{\epsilon }_j }(z)\bar{\phi }_{aa+\bar{\epsilon }_j }(z) & = &
I_n ,
\end{eqnarray}
This follows from the definitions of
$\phi_{ab}(z)$ and $\psi_{ab}(z)$ and the following lemmas:

{}~

{\bf Lemma 3.4}
 \cite{RT}~~~~
{\it If $f$ is an entire function not identically zero
which satisfies }
\begin{equation}
\left\{ \begin{array}{rcl}
f(z+\tau ) & = & \exp [-2\pi \sqrt{-1}(A_1 +A_2 z)]f(z), \\
f(z+1) & = & \exp (-2\pi \sqrt{-1}B)f(z), \end{array} \right.
\end{equation}
{\it then necessarily $A_2$ is a positive integer,
and $f$ has $A_2 $ zeros $u_1 , \cdots , u_{A_2 }$ in $E_{\tau }$ with }
\begin{equation}
\sum_{i=1}^{A_2 } u_i =\frac{A_2 }{2}+B\tau -A_1 , ~~~~
\mbox{mod~} \Lambda _{\tau }.
\end{equation}
It can be easily shown by using an elementary complex analysis.

{}~

{\bf Lemma 3.5}~~~~
\begin{equation}
\det \left( \begin{array}{ccc}
\theta ^{(0)}(nz_0 ) & \cdots & \theta ^{(0)}(nz_{n-1} ) \\
\vdots & ~ & \vdots \\
\theta ^{(n-1)}(nz_0 ) & \cdots & \theta ^{(n-1)}(nz_{n-1} ) \end{array}
\right) =C~h(\sum_{j=0}^{n-1} z_j -(n-1)/2)
\prod_{i>k} h(z_i -z_k ), \label{FZ}
\end{equation}
{\it where, $C$ is a constant defined by (\ref{C}). }

[Proof]~~Let $f(z_0 )$ stand for the difference of
the l.h.s and the r.h.s. of (\ref{FZ}).
Note that $f(z_0 )$ is an entire function of $z_0 $.
We have the following quasi-periodicities
\begin{equation}
\left\{ \begin{array}{rcl}
f(z_0 +\tau ) & = & \exp [-2\pi \sqrt{-1}(n\tau /2+1/2 +nz_0 )]f(z_0 ), \\
f(z_0 +1) & = & (-1)^n f(z_0 ). \end{array} \right.
\end{equation}
{}From Lemma 3.4 $f(z_0 )$ has $n$ zeros in $E_{\tau }$
and the sum of these zeros is $(n-1)/2 ~~(\mbox{mod } \Lambda _{\tau})$.
Clearly $f(z_0 )=0$ at $z_0 =z_1 , \cdots , z_{n-1}$ and hence
the last zero is located at
$z_0 =-(z_1 +\cdots +z_{n-1})+(n-1)/2 $.
By repeating the same argument for other variables
$z_1 , \cdots , z_{n-1}$, we obtain (\ref{FZ})
with an appropriate constant $C$.

Let us determine the constant $C$.
Put $z_i =(1/2+i\tau )/n$ for $0\leq i<n$.
By using the transformation property
\begin{equation}
\theta ^{(j)}(z+\tau )=\theta ^{(j-1)}(z)
\exp \frac{2\pi \sqrt{-1}}{n}(z+\frac{1}{2}+\frac{\tau }{2}),
\end{equation}
and some linear algebra, we obtain (\ref{C}). $\Box$

{}~

Therefore $\bar{\phi }_{ab}(z)$'s are right intertwining vector
(see Figure 3b):

{}~

{\bf Theorem 3.6}
\begin{equation}
\bar{\phi }_{ab}(z_2 )\otimes
\bar{\phi }_{bc}(z_1 )\check{R}(z_1 -z_2 ) =
\sum_{d} W_{z_1 -z_2 }
\left( \begin{array}{cc} a & b \\ d & c \end{array} \right)
\bar{\phi }_{ad}(z_1 )\otimes
\bar{\phi }_{dc}(z_2 ). \label{QF}
\end{equation}
[Proof]~~Multiply (\ref{JMO}) by
$\bar{\phi }_{ab'}(z_2 )\otimes \bar{\phi }_{b'c'}(z_1 )$
from the left and
by $\bar{\phi }_{ad}(z_1 )\otimes \bar{\phi }_{dc}(z_2 )$
from the right,
and sum over $c$ and $d$.
Then we obtain (\ref{QF}) from Proposition 3.3. $\Box$

{}~

The expression of the right intertwining vectors (\ref{phibar})
was first introduced in \cite{INS}.
We learn from Hasegawa that he also obtained the similar results \cite{H}.
It follows from Theorem 2.2 and 3.6 that
there exists a factorized representation of $\frak F _n $:

{}~

{\bf Theorem 3.7}~~~~ $L_{ab}(z)=\phi_{ab}(z)\bar{\phi }_{ab}(z)$
{\it solves the eq.(\ref{QIM}). }

{}~

We call this representation a factorized representation.

{}~

\section{$A^{(1)}_{n-1}$ Broken \protect\( {\protect\Bbb Z}_N \protect\) Model}
4.1~~{\it General Theory of Construction of a Solvable Model}~~~~
In this subsection we explain how to solve the following equation
\begin{equation}
S(z_1 ,z_2 )L(z_1 )\otimes L(z_2 )=
L(z_2 )\otimes L(z_1 )S(z_1 ,z_2 ), \label{SLL}
\end{equation}
where the symbol of the sum
representing the product of matrices
in $V=\bc ^n $ is omitted.
To construct a solution of (\ref{SLL}) we introduce (see Figure 4)
\begin{equation}
\bar{\psi }_{a-\bar{\epsilon }_i a}(z)=
{}~^{t}\phi _{a a+\bar{\epsilon }_i}(-z),~~~~
\psi _{a-\bar{\epsilon }_i a}(z)
=~^{t}\bar{\phi }_{a a+\bar{\epsilon }_i}(-z).
\end{equation}
It is clear that
the orthogonal relations hold for
$\psi _{ab}(z)$ and $\bar{\psi }_{ab}(z)$

{}~

{\bf Proposition 4.1}~~
\begin{eqnarray}
\bar{\psi }_{a-\bar{\epsilon }_i a}(z)\psi _{a-\bar{\epsilon }_k a}(z) & = &
\delta_{ik}, \\
\sum_{j=0}^{n-1}
\psi _{a-\bar{\epsilon }_j a}(z)\bar{\psi }_{a-\bar{\epsilon }_j a}(z) & = &
I_n .
\end{eqnarray}

{}~

By the factorization property (\ref{SLL}) reduces to
the smaller equation (see Figure 5, 6a)
\cite{HY}\cite{SLN}
\begin{equation}
\overline{W}_{ab}(z,z')\cdot (\bar{\phi }_{ad}(z')\phi _{bc}(z))=
(\bar{\psi }_{ad}(z)\psi _{bc}(z'))\cdot
\overline{W}_{dc}(z,z'). \label{small}
\end{equation}
Here $a,b,c,d \in \frak h (a_0 ) $.
On the assumption of the existence of $\overline{W}_{ab}(z,z')$
we set
\begin{equation}
S(z_1 ,z' _1 ;z_2 ,z_2 ')
:=\sum
S\left( \begin{array}{cc} a & b \\ d & c \end{array} \right.
\left| \begin{array}{cc} z_1 & z_2 \\ z_2 ' & z_1 ' \end{array} \right)
E_{ab} \otimes E_{dc} ,
\end{equation}
where,
\begin{equation}
S\left( \begin{array}{cc} a & b \\ d & c \end{array} \right.
\left| \begin{array}{cc} z_1 & z_2 \\ z_2 ' & z_1 ' \end{array} \right)
=\frac{\overline{W}_{ab}(z_1 ,z_2 )\overline{W}_{bc}(z_2 ,z_1 ')
\overline{W}_{cd}(z_1 ',z_2 ')}{\overline{W}_{ad}(z_1 ,z_2 ')}.
\end{equation}
and $(E_{ab})$ is the standard basis in $End(\frak h (a_0 ))$.

{}~

{\bf Theorem 4.2}~~{\it If we put
$L_{ab}(z,z'):=\phi_{ab}(z)\bar{\phi }_{ab}(z')$, then}
\begin{equation}
S(z_1 ,z' _1 ;z_2 ,z_2 ')
L(z_1 ,z_1 ')\otimes L(z_2 , z_2' )
=L(z_2 , z_2' )\otimes L(z_1 ,z_1 ')
S(z_1 ,z' _1 ;z_2 ,z_2 '). \label{THM4.4}
\end{equation}
[Proof]~~
First we notice the following relations (see Figure 6b,c):
\begin{eqnarray}
\sum_{a}
\overline{W}_{ab}(z,z')\cdot (\psi _{ad}(z)\bar{\phi }_{ad}(z')) & = &
\sum_{c} (\psi _{bc}(z')\bar{\phi }_{bc}(z))\cdot \overline{W}_{dc}(z,z').
\label{W1} \\
\sum_{b}
\overline{W}_{ab}(z,z')\cdot (\phi _{bc}(z)\bar{\psi }_{bc}(z')) & = &
\sum_{d}
(\phi _{ad}(z')\bar{\psi }_{ad}(z))\cdot \overline{W}_{dc}(z,z'). \label{W2}
\end{eqnarray}
These are due to Propositions 3.3 and 4.1.
Multiplying (\ref{small}) by $\psi _{ad}(z)$ from the left,
and by $\bar{\phi }_{bc}(z)$ from the right and summing over
$a$ and $c$,
we obtain (\ref{W1}).
Eq. (\ref{W2}) is obtained similarly.

The claim of this theorem is equivalent to
\begin{eqnarray}
&&\sum_{bc} S\left( \begin{array}{cc} a & b \\ d & c \end{array} \right.
\left| \begin{array}{cc} z_1 & z_2 \\ z_2 ' & z_1 ' \end{array} \right)
L_{bf}(z_1 ,z_1 ')\otimes L_{ce}(z_2 , z_2' ) \nonumber \\
&&=\sum_{bc} L_{ab}(z_2 , z_2' )\otimes L_{dc}(z_1 ,z_1 ')
S\left( \begin{array}{cc} b & f \\ c & e \end{array} \right.
\left| \begin{array}{cc} z_1 & z_2 \\ z_2 ' & z_1 ' \end{array} \right).
\label{fac}
\end{eqnarray}
Since $S$-matrix factorizes into four $\overline{W}$'s and
$L$-operator factorizes into $\phi $ and $\bar{\phi }$,
(\ref{fac}) can be shown by successive application of
(\ref{small}), (\ref{W1}), and (\ref{W2}). $\Box$

{}~

{\bf Corollary 4.3}~~{\it If we put $z_1 =z_1 ', z_2 =z_2 '$,
then $S(z_1 ,z_2 )=S(z_1 ,z_1 ;z_2 ,z_2 )$ solves eq.(\ref{SLL}).}

{}~

4.2~~{\it New Model}~~~~
Let us focus on eq.(\ref{small}).
Set $d=a+\bar{\epsilon }_i , c=b+\bar{\epsilon }_k $.
Then $\bar{\phi }_{ad}(z)\phi _{bc}(z')$ and
$\bar{\psi }_{ad}(z')\psi _{bc}(z)$ are expressed in terms of the
determinant in End($V$).
{}From Lemma 3.5 we have

{}~

{\bf Theorem 4.4}~~~~
{\it If we set }
$\overline{W}_{ab}(z,z')=
\sqrt{\Delta (a)}\sqrt{\Delta (b)}\overline{w}_{ab}(z,z')$,
{\it where, }
\begin{equation}
\Delta (a):=\prod_{l<m} h(a_{lm}w) ,
\end{equation}
{\it then we have }
\begin{equation}
\frac{\overline{w}_{a+\bar{\epsilon }_i b+\bar{\epsilon }_k}(z,z')}
{\overline{w}_{ab}(z,z')}
=\frac
{\prod_{m=0}^{i-1} (h(u+(\bar{a}_m -\bar{b}_k )w)
\prod_{m=i+1}^{n-1} (h(-u-(\bar{a}_m -\bar{b}_k )w)}
{\prod_{m=0}^{k-1} (h(-u-(\bar{a}_i -\bar{b}_m +1)w)
\prod_{m=k+1}^{n-1} (h(u+(\bar{a}_i -\bar{b}_m +1)w)}
\end{equation}
{\it where, } $u=(z-z' )/n$.

{}~

In what follows we denote
$\overline{W}_{ab}(z,z')$ by $\overline{W}_{ab}(u)$.
It is easy to check the self-consistency of
the recursion formula of $\overline{W}_{ab}$'s such as
\begin{equation}
\frac{\overline{W}_{a+\bar{\epsilon }_i b+\bar{\epsilon }_k}(u)}
{\overline{W}_{ab}(u)}
\frac{\overline{W}_{a+\bar{\epsilon }_i +\bar{\epsilon }_j
b+\bar{\epsilon }_k +\bar{\epsilon }_l}(u)}
{\overline{W}_{a+\bar{\epsilon }_i b+\bar{\epsilon }_k}(u)}
=\frac{\overline{W}_{a+\bar{\epsilon }_j b+\bar{\epsilon }_k}(u)}
{\overline{W}_{ab}(u)}
\frac{\overline{W}_{a+\bar{\epsilon }_i +\bar{\epsilon }_j
b+\bar{\epsilon }_k +\bar{\epsilon }_l}(u)}
{\overline{W}_{a+\bar{\epsilon }_j b+\bar{\epsilon }_k}(u)},
\end{equation}
and
\begin{equation}
\frac{\overline{W}_{a(n) b(n)}(u)}
{\overline{W}_{ab}(u)}=
\prod_{j=0}^{n-1}
\frac{\overline{W}_{a(j+1) b(j+1)}(u)}
{\overline{W}_{a(j) b(j)}(u)} =1,
\end{equation}
where, $a(j):=a+\bar{\epsilon }_0 +\cdots +\bar{\epsilon }_{j-1} $.

Now we impose the cyclic condition
\begin{equation}
\overline{W}_{a+N\bar{\epsilon }_i b+N\bar{\epsilon }_k}(u)
=\overline{W}_{ab}(u).
\end{equation}
In other words,
the local states are regarded as
\begin{equation}
a\in P_N /N{\cal A}.
\end{equation}
Then we have $w=1/N$.
This gives a generalization of the broken $\bz _N$ model
\cite{KM}\cite{1P}.
Obviously the number of independent local states is $N^{n-1}$.
In what follows we assume that $n$ and $N$ are coprime
for simplicity.

We list the properties of this model.

(1)~Initial condition~~~Under an appropriate normalization, the following
holds:
\begin{equation}
\overline{W}_{ab}(0)=\delta _{ab}.
\end{equation}
[Proof]~~
If we set $z=z'$ and $c=d$ in (\ref{small}), then we have
\begin{equation}
\overline{W}_{ab}(0)\cdot (\bar{\phi }_{ac}(z)\phi _{bc}(z))=
\delta_{ab} \overline{W}_{cc}(0). \label{initial}
\end{equation}
Since $\bar{\phi }_{ac}(z)\phi _{bc}(z)\neq 0$ for generic $z$,
we get $\overline{W}_{ab}(0)=0$ unless $a=b$.
By putting $a=b$ in (\ref{initial}) we obtain
$\overline{W}_{aa}(0)=\overline{W}_{cc}(0)$. $\Box$

This property implies $S(z,z';z,z')=I\otimes I$.

(2)~Inversion relation~~~With a certain function $U(u)$, the
following holds:
\begin{equation}
\sum_{c} \overline{W}_{ac}(u)\overline{W}_{cb}(-u)
=U(u)\delta _{ab} \label{unitary}.
\end{equation}
[Proof]~~~~Multiply $\overline{W}_{ce}(-u)$ to (\ref{W2}) and
sum over $c$. Then we have
\begin{equation}
\sum_{c} V_{ac}(u)\cdot (\phi _{ce}(z')\bar{\psi}_{ce}(z))=
\sum_{d} (\phi _{ad}(z')\bar{\psi}_{ad}(z))\cdot V_{de}(u),  \label{vpp}
\end{equation}
where,
$V_{ab}(u):=\sum_{c} \overline{W}_{ac}(u) \overline{W}_{cb}(-u)$.
Multiplying (\ref{vpp}) by $\bar{\phi}_{ad'}(z')$ from the left,
by $\psi _{c'e}(z)$ from the right and
performing the product of matrices
in $V=\bc ^n $, we find
\begin{equation}
V_{ac'}(u)\cdot (\bar{\phi}_{ad'}(z')\phi _{ce'}(z'))=
(\bar{\psi}_{ad'}(z)\psi _{c'e}(z))\cdot V_{d'e}(u). \label{V}
\end{equation}
By applying the similar argument in (1) to (\ref{V})
we have (\ref{unitary}) with
the scalar factor $U(u)=V_{aa}(u)$.
$\Box$

This property implies
$S(z_1 ,z'_1 ;z_2 ,z'_2 )S(z_2 ,z'_2 ;z_1 ,z'_1 )=
U(z_1 -z_2)U(z'_2 -z'_1 )I\otimes I.$
We give the explicit expression of $U(u)$
for $n=3$ and $N=2$ in Appendix.

{}~

4.3~~{\it Yang-Baxter Equation}~~~~
Now we prove YBE for $S(z_1 ,z' _1 ,z_2 ,z' _2 )$
\begin{eqnarray}
&&S^{(23)}(z_1 ,z' _1 ;z_2 ,z' _2 )
S^{(12)}(z_1 ,z' _1 ;z_3 ,z' _3 )
S^{(23)}(z_2 ,z' _2 ;z_3 ,z' _3 ) \nonumber \\
&&=S^{(12)}(z_2 ,z' _2 ;z_3 ,z' _3 )
S^{(23)}(z_1 ,z' _1 ;z_3 ,z' _3 )
S^{(12)}(z_1 ,z' _1 ;z_2 ,z' _2 ), \label{SSS}
\end{eqnarray}
where,
$S^{(12)}(z_1 ,z' _1 ;z_2 ,z' _2 )=
S(z_1 ,z' _1 ;z_2 ,z' _2 )\otimes I$ and
$S^{(23)}(z_1 ,z' _1 ;z_2 ,z' _2 )=
I\otimes
S^{(23)}(z_1 ,z' _1 ;z_2 ,z' _2 )$.

Our proof is parallel to the ones given in Refs.\cite{Cycl3}.
First we prove the following proposition.

{}~

{\bf Proposition 4.5}~~{\it
Fix a positive integer $m$ and consider $m$-fold tensor product of
$L$-operators}
\begin{equation}
Q_m (z_1 ,z'_1 ; \cdots ; z_m , z'_m ):=
L(z_1 ,z'_1 )\otimes \cdots \otimes L(z_m ,z'_m ).
\end{equation}
{\it Then for generic $z_i $'s the only commutant of
$Q_m $ is the identity up to a constant. }

[Proof]~~Define $\chi _{ab}(z)$ by
\begin{equation}
\chi _{ab}(z) =\frac{h(z-(n-1)/2)\phi _{ab}(z)}{h(z+(n-1)(w-1/2))},
\end{equation}
then we have
\begin{eqnarray}
\bar{\phi }_{a-\bar{\epsilon }_i a}(z)
\chi_{a-\bar{\epsilon }_k a}(z) & = &
\delta_{ik}, \label{chi} \\
\sum_{j=0}^{n-1}
\chi_{a-\bar{\epsilon }_j a}(z)
\bar{\phi }_{a-\bar{\epsilon }_j a}(z) & = &
I_n .
\end{eqnarray}

{\it step 1}~~Suppose that $P(z,z')$ commutes with $Q_1 $
\begin{equation}
\sum_{d} P_{a}^{d}(z,z') \phi_{dc}(z) \bar{\phi }_{dc}(z') =
\sum_{b} \phi_{ab}(z) \bar{\phi }_{ab}(z') P_{b}^{c}(z,z'). \label{m1}
\end{equation}
Here and hereafter we omit the symbol of the sum
representing the multiplication of matrices in $V$.
Multiply (\ref{m1}) by $\bar{\phi }_{ac'}(z)$ from the left,
and by $\chi _{d' c}(z')$ from the right.
Then we have after setting $a=d' $
\begin{equation}
P_{a}^{a}(z,z') \delta ^{c}_{c'}=
D_1 (a,c,c';z')P_{c'}^{c}(z,z'), \label{m12}
\end{equation}
where,
\begin{equation}
D_1 (a,c,c';z')=(\bar{\phi }_{ac'}(z')\chi _{ac}(z')).
\end{equation}
Since
$D_1 (a,c,c';z')\neq 0$ for generic $z'$, we obtain
$\displaystyle P^{c}_{c'}\propto \delta ^{c}_{c'}.$
By putting $c'=c$ in (\ref{m12}) we further have $P_{a}^{a}=P_{c}^{c}$
and therefore $P(z,z')=\rho (z,z')I$.

{\it step 2}~~Assume that
$P(z_1 ,z'_1 ;\cdots ; z_m ,z'_m )$ commutes with $Q_m $ with $m>1$
\begin{equation}
\sum_{d_1 , \cdots , d_m } P_{a_1 , \cdots , a_m }^{d_1 , \cdots , d_m }
Q_{d_1 , \cdots , d_m }^{c_1 , \cdots , c_m }
=\sum_{b_1 , \cdots , b_m }
Q_{a_1 , \cdots , a_m }^{b_1 , \cdots , b_m }
P_{b_1 , \cdots , b_m }^{c_1 , \cdots , c_m }. \label{mg}
\end{equation}
Multiply (\ref{mg}) by  $\bar{\phi }_{a_1 ,c'_1}(z_1)$ from the left,
by $\chi _{d'_m c_m}(z'_m)$ from the right,
and by the product of scalars
$\displaystyle \prod_{i=1}^{m-1} \bar{\phi }_{d'_{i+1}c_{i+1}}(z_{i+1})
\chi _{d'_i c_i }(z'_i )$
in $V$.
Sum over $c'_2 ,\cdots , c'_m$,
then we have after setting $a_1=d'_1 $
\begin{equation}
P_{a_1 , a_2 , \cdots , a_m }^{a_1 , d'_2 , \cdots , d'_m }
\delta _{c'_1 }^{c_1 }
=\sum_{\scriptstyle{\begin{array}{l} b_2 ,\cdots ,b_m \\
c_2 ,\cdots ,c_m \end{array}}}
D_m (a,b,c,d';z,z')
P_{c'_1 , b_2 ,\cdots b_m }^{c_1 ,c_2 , \cdots ,c_m},
\end{equation}
where,
\begin{eqnarray}
D_m (a,b,c,d';z,z')
& = & \prod _{i=1}^{m-1}
(\bar{\phi }_{a_{i}b_{i}}(z_{i})\phi _{a_{i+1} b_{i+1}}(z'_{i+1}))
(\bar{\phi }_{d'_{i+1}c_{i+1}}(z_{i+1})\chi _{d'_i c_i }(z'_i ))\times
\nonumber \\
&&\times \bar{\phi }_{a_{m}b_{m}}(z_{m})\chi _{d'_m c_m }(z'_m ).
\end{eqnarray}
Since
$D_m (a,b,c,d';z,z')$ is non-degenerate for generic case,
we obtain
$P_{c'_1 , b_2 , \cdots , b_m }^{c_1 , c_2 , \cdots , c_m }
\displaystyle \propto \delta ^{c_1 }_{c'_1 }$.
By putting $a_1 =d'_1 , b_1 = c'_1 ,$ one can
confirm that
$P_{c_1 , b_2 , \cdots , b_m }^{c_1 , c_2 , \cdots , c_m }$ does
not depend on $c_1$.

{\it step 3}~~
The claim of this proposition follows from {\it step 1} and {\it step 2}
by induction. $\Box$

{}~

{\bf Theorem 4.6}~~{\it $S(z_1 ,z'_1 ;z_2 ,z'_2 )$ satisfies
YBE (\ref{SSS}).}

[Proof]~~Owing to the intertwining property (\ref{THM4.4}),
$R_{R}^{-1}S_{L}$, where $S_{L}$ and $S_{R}$ denote
the l.h.s. and the r.h.s. of (\ref{SSS}) respectively,
commutes with
$L(z_1 ,z'_1 )\otimes L(z_2 ,z'_2 )\otimes L(z_2 ,z'_3 )$.
Therefore from Proposition 4.6, we have $R_{R}^{-1}S_{L}=
\rho (z_1 ,z'_1 ;z_2 ,z'_2 ;z_3 ,z'_3 )I\otimes I\otimes I.$
By comparing the determinant of both sides
$\rho $
must be a root of unity and hence a constant in
$z_1 ,z'_1 ,z_2 ,z'_2 ,z_3 ,z'_3 .$
In particular by putting
$z_1 =z_2 =z_3, z'_1 =z'_2 =z'_3 $ we obtain
$\rho =1$. $\Box$

{}~

\section{Concluding Remark}
In this paper we constructed a factorized representation
of $\frak F _n $, the $\frak g \frak l _n $-Sklyanin algebra.
In our construction the concept of the
vertex-face correspondence is crucial;
i.e. the $L$-operator factorizes elementwise to the product
of the left and right intertwining vectors
between the $\bz _n \otimes \bz _n $-symmetric model
and the cyclic $A^{(1)}_{n-1}$ model.
We have further obtained a new solution of YBE
by utilizing the factorized representation.
The model obtained in this way gives an
$\frak s \frak l _n $-generalization
of the broken $\bz _N $ model.

We know two different
$N$-state generalizations
of the Ising model.
One is the chiral Potts model \cite{CP} which keeps the
$\bz _N$-invariance at any temperature
but is no longer parametrized on elliptic curves.
The other is the broken $\bz _N$ model \cite{KM}
whose Boltzmann weights are expressed in terms of the elliptic functions
but the $\bz _N$ symmetry is restored only at the critical point.
Both of these models give two different off-critical extensions
of the self-dual $\bz _N$ model \cite{FZ},
whose critical behavior is described by
the $\frak s \frak l _2$-parafermion model \cite{Para}.

It is expected that an $\frak s \frak l _n$-generalization
of both the chiral Potts model \cite{SLN}\cite{Cycl3}
and the broken $\bz _N $ model
are described by the $\frak s \frak l _n$-parafermion models at
the scaling limit.
To clarify this claim we must investigate the
critical behavior of
the one-point function of these model.
We shall discuss this problem in a subsequent paper.

Finally we would like to mention
the connection with the solvable model in three dimensions.
In a recent work
Bazhanov and Baxter \cite{BB} pointed out the connection
between the $\frak s \frak l _n $-chiral Potts model and
the solvable three dimensional model of Zamolodchikov
\cite{3d}\cite{3dB}.
It is very likely that our new model is also related with
the model in three dimensions.

{}~

\section*{Acknowledgement}
The authors thank
T. Deguchi, S. Iso, T. Nakanishi and J. Shiraishi
for variable discussions.
They also thank
T. Eguchi for continuous encouragement.
This work is supported in part by
the Grant-in-Aid for Scientific Research from the Ministry of Education,
Science and Culture (No. 04-2297 and No. 04-2747).

{}~

\appendix
\section{Simple Case Study}
In this appendix, we show the explicit forms of the Boltzmann weights
of the $n=3$,$N=2$ case. Here $w=1/2$.

For convenience, we denote $2\Lambda_{0}$,
$\Lambda_{0}+\Lambda_{1}$,
$2\Lambda_{1}$,$\Lambda_{1}+\Lambda_{2}$,$2\Lambda_{2}$
and $\Lambda_{2}+\Lambda_{0}$
by $A$,$B$,$C$,$D$,$E$ and $F$, respectively. (See Figure 7.)

One can write down $\overline{W}$'s as
\begin{eqnarray}
\overline{W}_{AA}(u)&=&1, \\
\overline{W}_{BB}(u)=\overline{W}_{DD}(u)&=&\overline{W}_{FF}(u)=
\frac{h(1/6)^2 }{h(1/3)^2 }
 {h(u+1/3)h(u-1/3) \over
 h(u+1/6)h(u-1/6)}, \\
\overline{W}_{AB}(u)=\overline{W}_{BA}(u)&=&
-\overline{W}_{AD}=-\overline{W}_{DA}(u) \cr
&=&-\overline{W}_{AF}(u)=-\overline{W}_{FA}(u) \cr
&=&\sqrt{-1}{h(1/6) \over h(1/3)}
{h(u)h(u-1/3)h(u+1/3) \over
h(u+1/2)h(u-1/6)h(u+1/6)}, \label{im} \\
\overline{W}_{BD}(u)=\overline{W}_{DF}(u)&=&\overline{W}_{FB}(u)=
\frac{h(1/6)^2 }{h(1/3)^2 }
{h(u)h(u-1/3)
\over h(u+1/2)h(u+1/6)}, \\
\overline{W}_{BF}(u)=\overline{W}_{FD}(u)&=&\overline{W}_{DB}(u)=
-\frac{h(1/6)^2 }{h(1/3)^2 }
{h(u)h(u+1/3)
\over h(u+1/2)h(u-1/6)}.
\end{eqnarray}
One can easily see that
\begin{equation}
\overline{W}_{CC}(u)=\overline{W}_{AA}(u),~~~~
\overline{W}_{CB}(u)=\overline{W}_{AB}(u),
\end{equation}
etc., and hence the independent local states are $A,B,D,F$.
There exists an imaginary unit $\sqrt{-1}$ in (\ref{im}).
However,
the imaginary unit disappear in $S$-matrix because
(\ref{im})-type terms always appear in pairs
in $\displaystyle
S\left( \begin{array}{cc} a & b \\ d & c \end{array} \right.
\left| \begin{array}{cc} z_1 & z_2 \\ z_2 ' & z_1 ' \end{array} \right) $
for arbitrary $a,b,c,d \in \{ A,B,D,F \}$.

For the present case,
the initial condition and the inversion relation
can be easily checked and
the scalar factor $U(u)$ in (\ref{unitary}) becomes
\begin{equation}
U(u)=1-3\left[
{h(1/6) \over h(1/3)}
{h(u)h(u-1/3)h(u+1/3) \over
h(u+1/2)h(u-1/6)h(u+1/6)} \right]^{2}.
\end{equation}

\newpage
{\Large \bf Figure Captions}

\begin{description}
\item[Figure 1~] Boltzmann weight for the cyclic $A^{(1)}_{n-1}$ model.
\item[Figure 2a] Right intertwining vector.
\item[Figure 2b] Intertwining property for $\phi (z)$.
\item[Figure 3a] Left intertwining vector.
\item[Figure 3b] Intertwining property for $\bar{\phi }(z)$.
\item[Figure 4~] Other intertwining vectors.
\item[Figure 5~] One of four factors consisting of
the Boltzmann weight
for $A^{(1)}_{n-1}$ broken $\bz _N $
model.
\item[Figure 6~] Relations among $\overline{W}$'s and intertwining vectors.

\item[Figure 7~] Set of local states
of $A^{(1)}_{n-1}$ broken $\bz _N $ model
for $n=3, N=2$. Owing to the periodicity of $2{\cal A}$,
one must identify $A, C$, and $F$.
\end{description}
\end{document}